\documentclass[prl,twocolumn,showpacs,preprintnumbers,amsmath,amssymb]{revtex4}
\usepackage{graphicx}% Include figure files
\usepackage{dcolumn}% Align table columns on decimal point
\usepackage{bm}% bold math

\begin{document}

%\preprint{APS/123-QED}

\title{Fermi-liquid ground state in  n-type copper-oxide superconductor Pr$_{0.91}$LaCe$_{0.09}$CuO$_{4-y}$}% Force line breaks with \\

%\author{Guo-qing~Zheng}
% \altaffiliation[Present address : ]{Department of Physics, Faculty of Engineering, Tokushima University, Tokushima 770-8506, Japan}%
%\author{K.~Ishida}%
 %\altaffiliation[Present address : ]{Department of Physics, Graduate School of Science, Kyoto University, Kyoto 606-8502, Japan}%
%\author{S.~Kawasaki}%
%\author{T.~Mito}%
% \altaffiliation[Present address : ]{Department of Physics, Faculty of Science, Kobe University, Hyogo 657-8501, Japan}%

\author{Guo-qing~Zheng $^1$}%
\author{T.~Sato $^1$}%
\author{Y.~Kitaoka $^1$}%
\author{M.~Fujita $^2$}%
\author{K.~Yamada $^2$}%

\address{$^1$ Department of Physical Science, Graduate School of Engineering Science, Osaka University,  Osaka 560-8531, Japan}
\address{$^2$ Institute for Chemical Research, Kyoto University, Uji, Kyoto 610-0011, Japan}
%\affiliation{%
%\affiliation{

\date{\today}% It is always \today, today,

\begin{abstract}
We report nuclear magnetic resonance studies on the low-doped n-type copper-oxide Pr$_{0.91}$LaCe$_{0.09}$CuO$_{4-y}$ ($T_c$=24 K) in the superconducting state and in the normal state uncovered by the application of a strong magnetic field.  We  find that when the superconductivity is removed, the underlying ground state is the Fermi liquid state.  This result is  at variance with that inferred from previous thermal conductivity  measurement and  contrast with  that in  p-type copper-oxides with a similar doping level where high-$T_c$ superconductivity sets in within the pseudogap phase.  The data in the superconducting state are  consistent with the line-nodes gap model.
\end{abstract}

\maketitle

%\section{Introduction}
\sloppy
The underlying ground state from which superconductivity evolves is closely related to, and may even determine, the nature of the superconductivity. In conventional metals, BCS type of superconductivity  develops out of a ground state described by Landau's Fermi liquid theory in which electrons, even interact with each other, can be treated as dressed fermions called quasiparticles \cite{Landau}. By contrary, the normal state above the transition temperature ($T_c$) in the p-type (hole-doped) copper-oxide superconductors deviates  \cite{Timusk} from the Fermi liquid. One of the emerging pictures is that high-$T_c$ superconductivity evolves out of a new state of matter \cite{Anderson}. Meanwhile, the n-type (electron-doped) copper-oxide superconductors Re$_{2-x}$Ce$_x$CuO$_{4-y}$ (Re=Nd, Pr, Eu or Sm) \cite{Tokura} show a substantially lower $T_c$ than their p-type counterparts. It is therefore an outstanding important question of what is the difference in the underlying ground state between these two classes of materials. Early experiments found that the electrical conductivity in Re$_{2-x}$Ce$_x$CuO$_{4-y}$ (Re=Nd, Pr) is strikingly different from that in the p-type copper-oxides \cite{Takagi,Peng}, but it is recently suggested that the ground state in Pr$_{1.85}$Ce$_{0.15}$CuO$_{4-y}$ is a "spin-charge separated" state, as the case thought to occur in the p-type materials, where electrons have different capability to transport heat and charge \cite{Hill}. 

In this study,        we  suppress superconductivity in the n-type Pr$_{0.91}$LaCe$_{0.09}$CuO$_{4-y}$ ($T_c$=24 K) with a strong magnetic field and evaluate the normal state using $^{63}$Cu nuclear magnetic resonance (NMR). We find that the ground state hidden behind superconductivity is the Fermi liquid. When the superconductivity is removed, the spin lattice relaxation rate ($1/T_1$) was found to follow the $T_1T$=constant relation, known as Korringa law \cite{Korringa}, down to a very low temperature of $T$=0.2 K. The NMR data in the superconducting state, which were obtained for the first time in Re$_{2-x}$Ce$_{x}$CuO$_{4}$,  are rather consistent with line-nodes gap model.

      High quality single crystal of Pr$_{1-x}$Ce$_x$LaCuO$_{4-y}$ used in this study was grown by the traveling-solvent-floating-zone method  \cite{Fujita}. Here half of Pr was replaced by La that helps stabilizing crystallization to obtain large-size single crystals. More crucially, it helps eliminating the magnetic moment due to the rare earth element that has been the main cause for preventing the understanding of this class of superconductors \cite{Zheng4,Cooper}. Substituting $x$ tetravalent Ce for trivalent Pr adds $x$ electron to the CuO$_2$ plane. Upon doping Ce, superconductivity appears at $x$=0.09 with  $T_c$ as high as 26 K, then $T_c$ decreases monotonically with increasing $x$  \cite{Fujita}.   A crystal of 5$\times$2$\times$0.5 mm$^3$ of Pr$_{0.91}$LaCe$_{0.09}$CuO$_{4-y}$ was used for NMR and ac magnetic susceptibility measurements. Figure 1 shows the critical field necessary to destroy superconductivity, $H_{c2}$, determined from the ac-susceptibility measured at high frequency of $f$=175 MHz (left inset of Fig. 1). The value of $H_{c2}$($T$=0) for the magnetic field ($H$) applied perpendicular to the CuO$_2$ plane ($H \parallel$ c-axis) is estimated to be less than 10 T, indicating a substantially longer  
superconducting-coherence-length than that in the p-type cuprates. This low $H_{c2}$  makes it possible to access  the "normal" ground state by applying a laboratory magnetic field to suppress  the superconductivity.  Note that in typical p-type superconductors, the highest static field of $\sim$30 T using Bitter magnets can only reduce $T_c$ to its half value \cite{Zheng1,Zheng2}.

Taking full advantage of the large anisotropy of $H_{c2}$, we study the nature of the superconductivity by applying a magnetic field of 6.2 T along the CuO$_2 $ plane ($T_c$(6.2T)=19 K) and we explore the ground state by applying a field of 15.3 T perpendicular to the plane to suppress the superconductivity.    A single $^{63}$Cu NMR line was observed for both $H \parallel$ c-axis and $H \parallel$ a-axis. The full width at half maximum of the linewidth at $T$=300 K is $\sim$150 Oe ($\sim$300 Oe) at $H$=6.2 T (15.3 T) $\parallel$c-axis and $\sim$75 Oe at $H$=6.2 T $\parallel$ a-axis, which indicate that the nuclear quadrupolar frequency $\nu_Q$ is very small ($\nu_Q \leq$ 0.5 MHz, if any), as the case in Nd$_{1.85}$Ce$_{0.15}$CuO$_4$ \cite{Zheng4}. This is understood as due to electron doping into the Cu 3d-orbit whose hole is the main contribution to $\nu_Q$ \cite{Zheng5}. The  $1/T_1$ of $^{63}$Cu was measured by the saturation-recovery method. A small RF field was used in order to avoid possible heating by the RF pulse. The value of $1/T_1$ was determined from an excellent fitting of the nuclear magnetization $M(t)$ to $\frac{M(\infty)-M(t)}{M(\infty)}=0.1exp(-t/T_1)+0.9exp(-6t/T_1)$ \cite{Narath}.

\begin{figure}
\begin{center}
\includegraphics[scale=0.47]{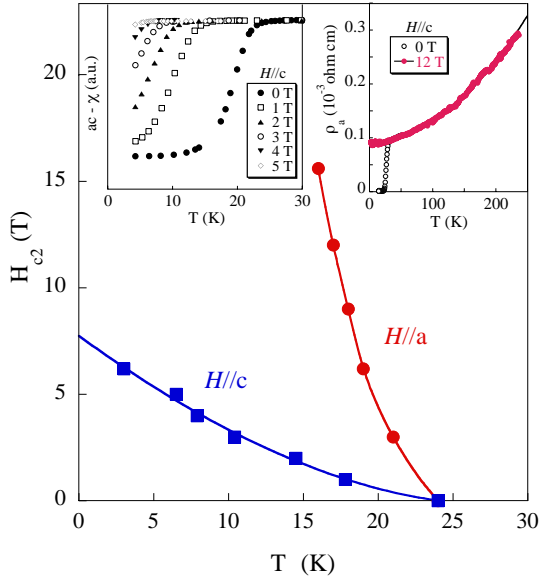}
\caption{The critical field of Pr$_{0.91}$LaCe$_{0.09}$CuO$_{4-y}$ as a function of temperature. 
%for the magnetic fields applied parallel and perpendicular to the conducting CuO$_2$ plane, respectively. 
The curves are guides to the eyes. The insets show the ac-susceptibility at various fields and the in-plane resistivity.}
\label{fig:1}
\end{center}
\end{figure}

The key results of this study were obtained by measuring $1/T_1$ in the zero-temperature limit "normal state" when the superconductivity is suppressed. A criterion for judging an electronic state being a Fermi liquid or a "strange metal" is to see whether or not $1/T_1$ obeys the Korringa law \cite{Korringa}. In a Fermi liquid, those quasiparticles that participate in the nuclear spin-lattice relaxation are the quasiparticles near $E_F$ whose population is $\sim k_BT$. Therefore $1/T_1$ is proportional to $T$ in such a state  (see below for a broader definition of the Korringa law). As seen in Fig. 2, for $H$ along the CuO$_2$ plane of Pr$_{0.91}$LaCe$_{0.09}$CuO$_{4-y}$, there appears a signature of $1/T_1$ becoming proportional to $T$ below a temperature  above $T_c$, but only in a narrow temperature range because superconductivity sets in (see Fig. 2,  circles) \cite{note1}. It is unclear whether or not this is in the middle of crossing over to a more exotic state at lower temperatures as seen in  p-type cuprate superconductors. We therefore applied a strong magnetic field of 15.3 T perpendicular to the CuO$_2$ plane to suppress the superconductivity and thereby to reveal the ground state hidden behind the superconductivity. There, it is seen that the $T_1T$=constant relation indeed holds, persisting down to a very low temperature $T$=0.2 K (also see the inset of Fig. 3). Namely, the hallmark of the Fermi liquid is unambiguously observed over more than two decades in temperature. 

\begin{figure}
\begin{center}
\includegraphics[scale=0.45]{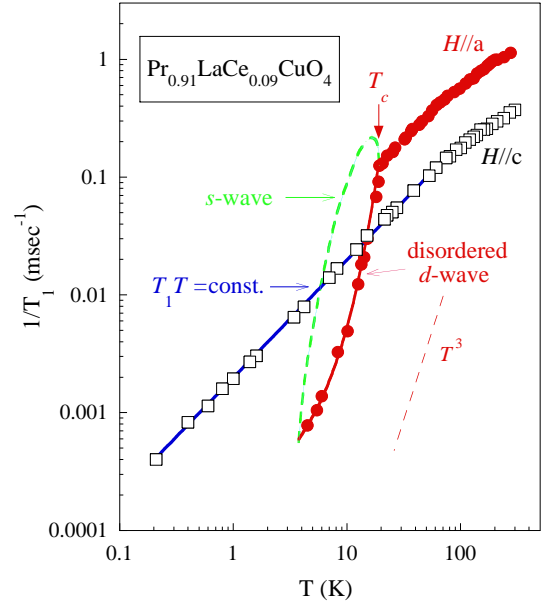}
\caption{Temperature dependence of the $^{63}$Cu nuclear spin-lattice relaxation rate, $1/T_1$, with the magnetic filed applied parallel (circles, $H$=6.2 T) and perpendicular (squares, $H$=15.3 T) to the CuO$_2$ plane, respectively. The solid curve is the calculated result for the disordered $d_{x^2-y^2}$ gap function \cite{Hotta}, $\Delta=\Delta_0 cos(2\theta)$, with 2$\Delta_0=3.8k_BT_c$  and a residual DOS at the Fermi level, $N_{res}=0.15N(E_F)$. The broken curve is the calculated result for a BCS s-wave gap, 2$\Delta_0=3.5k_BT_c$ and the ratio of $\Delta_0$ to the energy level breadth \cite{Hebel}, $r$=10.}
\label{fig:2}
\end{center}
\end{figure}

In order to examine in more detail how the electron-electron correlations evolve with temperature, it is instructive to plot the quantity of $1/T_1T$ as a function of $T$. In Fig. 3, one sees that upon lowering temperature from 300 K, $1/T_1T$ increases with decreasing $T$, but is saturated around $T$=70 K. The most important feature is that $1/T_1T$ becomes a constant below $T$=55 K which persists down to $T$=0.2 K (see the inset), as mentioned already. Note that $1/T_1T$ probes the imaginary part of the low-frequency ($\omega \rightarrow$0) dynamical susceptibility ($\chi(q,\omega)$) averaged over the momentum ($q$) space: $1/T_{1}T = \frac{3k_{B}}{4}\frac{1}{\mu_{B}^{2}\hbar^{2}}\sum_{q}A_{q}A_{-q}\frac{\chi^
{"}(q,\omega)}{\omega}$, where $A_q$ is the hyperfine coupling constant \cite{Moriya1}. For conventional metals described by the Fermi liquid theory, $\sum_q \chi"(q,\omega)=\pi\sum_{k,k'} \delta (E_k-E_{k'}-\hbar \omega)(f(E_{k'})-f(E_k))$,  thus one recovers the Korringa law of $1/T_1T=\frac{\pi}{\hbar}A^2N(E_F)^2k_B=\frac{4\pi k_B}{\hbar}(\frac{\gamma_n}{\gamma_e})^2K_s^2$, where  $\gamma_{e(n)}$ is the Gyromagnetic ratio of electron (nucleus) and $K_s$ is the spin Knight shift. By contrary, when $\chi(q)$ has a peak at the antiferromagnetic wave vector $Q=(\pi,\pi)$ as seen in the p-type cuprates, $1/T_1T$ becomes to be proportional to $\chi(Q)$ \cite{Millis,Moriya2}. The increase of $1/T_1T$ upon decreasing temperature in Pr$_{0.91}$LaCe$_{0.09}$CuO$_{4-y}$ can therefore be attributed to the increase of $\chi(Q)$, namely, to the development of antiferromagnetic spin correlations. But the  increase of $1/T_1T$ is weak,  resembling that in high-doped (overdoped) p-type materials \cite{Zheng2}, which may be due to the electron-doping into the Cu-3d orbit that reduces the size of the spin moment. More importantly, for the low-doped p-type copper-oxides with carrier density usually less than 0.2, with further lowering $T$, $1/T_1T$ starts to decrease at a temperature $T^*$ that is far above $T_c \sim$100 K. This phenomenon of the loss of low-energy spectral weight or DOS is ascribed to be due to a  pseudogap  opening \cite{Timusk}, which  has been a subject of intensive studies over the last decade.  However, the pseudogap behavior is not seen in $1/T_1T$  in Pr$_{0.91}$LaCe$_{0.09}$CuO$_{4-y}$ even though it is low-doped. Its low-$T$ ($T\leq$ 55 K) spin dynamics is renormalized to well conform to the prediction for the Fermi liquid that persists as the ground state when the superconductivity is removed.

\begin{figure}
\begin{center}
\includegraphics[scale=0.45]{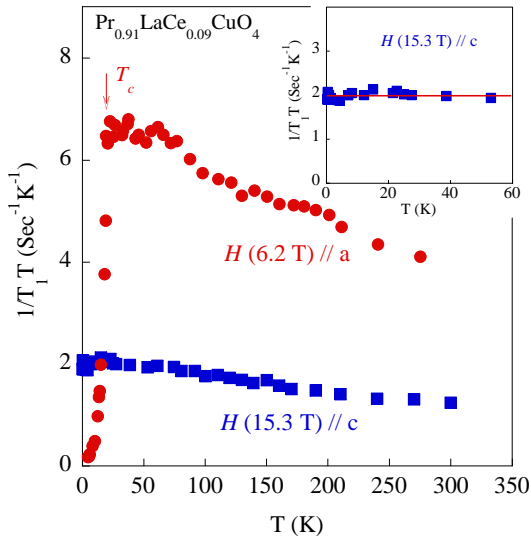}
\caption{The quantity of $1/T_1T$ as a function of temperature in Pr$_{0.91}$LaCe$_{0.09}$CuO$_{4-y}$. The inset  emphasized the low temperature part for $H\parallel c$-axis.}
\label{fig:3}
\end{center}
\end{figure}

That  the present compound is a low carrier-doped sample  is  supported by the Knight shift result.   The spin susceptibility $\chi (q=0)$ of the present sample  shares a property commonly seen in low-doped p-type cuprate superconductors \cite{Zheng3}, namely, $\chi (q=0)$ decreases with lowering $T$, although its $T_c$ is the highest among its class.  In Fig. 4, the Knight shift, which is a sum of the orbital contribution $K_{orb}$ and the spin contribution $K_s=A_{hf}\chi (q=0)$,  are shown for both $a$-direction ($K_a$) and $c$-direction ($K_c$). $K_c$ is less $T$-dependent, probably because of smaller hyperfine interaction $A_{hf}$ in this direction, namely, $K_c$ is predominantly orbital origin. As seen in the lower panel of Fig. 4 and the inset to it, the decrease of $K_a$ as $T$ is reduced ceases below $T\sim$55 K, namely, $K_a$ becomes $T$-independent within the experimental error below $T\sim$55 K, until superconductivity sets in. We have applied a high field of 28 T along the $a$-direction to reduce the $T_c$ to $\sim$ 10 K and confirmed that  the $T$-independence of $K_a$ persists  down to 10 K. Thus, the  criterion for the Fermi liquid in a broader sense, the requirement of $T_1TK_s^2$=constant, is also fulfilled  
for $T\leq$ 55 K. 
Taking $K_{orb,a}$=0.155\% from Fig. 4 \cite{Note}, we obtain $T_1TK_s^2$=7.5$\times$10$^{-8}$Sec$\cdot$K, which is smaller by a factor of 50 than the value for non-interacting electrons of 3.75$\times$10$^{-6}$Sec$\cdot$K. The difference primarily arises from the enhancement of $1/T_1T$ due to the antiferromagnetic spin correlation. Hence, the low-$T$ normal state is a {\it strongly-correlated Landau Fermi liquid}. The in-plane resistivity $\rho_{a}$, as shown in the right inset of Fig. 1,  can be fitted to $\rho_{a}=\rho_{0}+AT^2$ 
with $\rho_{0}$=92$\mu\Omega$cm and $A$=3.75$\times$10$^{-3}\mu\Omega$cmK$^{-2}$, which is consistent with such Fermi liquid state but inconsistent with other liquids \cite{Varma}. Our conclusion is different from that inferred from a recent low-$T$ thermal conductivity measurement that suggested a break-down of the Fermi liquid theory in the n-type Pr$_{1.85}$Ce$_{0.15}$CuO$_4$ \cite{Hill}. Also note that in  p-type cuprates, the pseudogap persists in the $H$-induced normal state below $T_c$($H$=0)  \cite{Zheng1,Ando}.   Thus, our results suggest that the physics of the ground state in the n-type copper-oxide superconductors differs  from that in the p-type counterparts, which may  be responsible for the strikingly different $T_c$ they generated. 

\begin{figure}
\begin{center}
\includegraphics[scale=0.45]{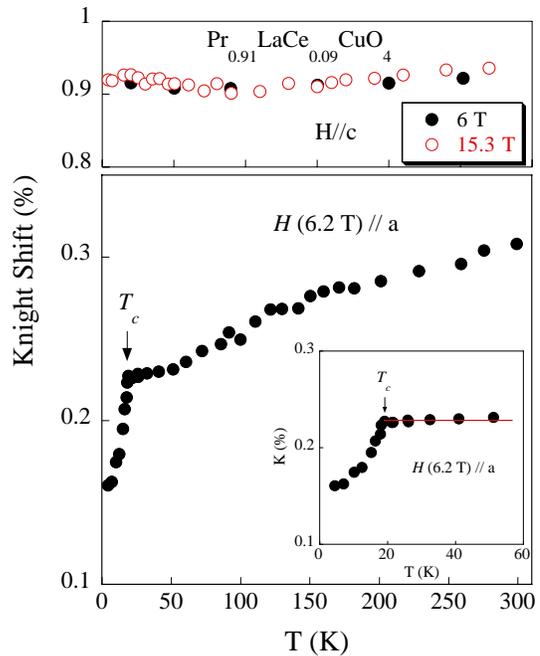}
\caption{Temperature dependence of the $^{63}$Cu Knight shift in Pr$_{0.91}$LaCe$_{0.09}$CuO$_{4-y}$. The inset to the lower panel emphasizes the data below $T$=60 K.}
\label{fig:4}
\end{center}
\end{figure}

Finally, we discuss the  nature of the superconducting gap. As seen in  Fig. 2, $1/T_1$ in the superconducting (SC) state for $H$ parallel to the CuO$_2$-plane ($H \parallel a$-axis) is reduced sharply just below $T_c$, and decreases in proportion to $T^n$ upon further lowering $T$, with the exponent $n=3$ for 6 K$\leq T \leq$ 13 K. As elaborated below, this is consistent with the existence of line-nodes  in the SC order parameter. 
In terms of density of states (DOS), $T_{1s}$ in the SC state is expressed as $\frac{T_{1N}}{T_{1s}} = \frac{2}{k_BT}\int \int   (1+\frac{\Delta^2}{E E^{'}})N_{s}(E) N_{s}(E^{'}) f(E)(1-f(E^{'}))\delta(E-E^{'})dEdE^{'}$,
where $N_{s}(E)$ is the DOS in the SC state,
$f(E)$ is the Fermi function, $\Delta$ is the energy gap. 
For an  s-wave gap, $1/T_1$ should show a peak just below $T_c$ due to the  divergence of $N_s$  at $E=\Delta$, and then decrease as $\sim exp(-\Delta/k_BT)$.  The broken curve in Fig. 2 is the  calculated $T$-dependence of $1/T_1$ for the simple BCS gap which was applied previously to several n-type cuprates \cite{Huang,Stadlober,Skinta,Chen}. By contrary, for a line-nodes gap such as d-wave gap, the peak just below $T_c$ is removed \cite{Mac}, and a $N_s(E)\propto E$ at low $E$ results in the $1/T_1 \propto T^3$ behavior at {\it low temperatures}. In the present case, the deviation of 
$1/T_1$ from  $T^3$  at  $T\leq$ 6 K is explained as due to  the presence of impurity (disorder) scattering  which     brings about a  residual DOS  ($N_{res}$) at the Fermi level ($E_{F}$) in the case of linenodes gap \cite{Hotta,Ferenbacher}. Note that an anisotropic (or extended) s-wave gap model can not explain the $T$-linear behavior of $1/T_1$ at low $T$ \cite{Ferenbacher}. To demonstrate the result of line-nodes gap, we applied 
 the  $d_{x^2-y^2}$ gap model, $\Delta(\theta)=\Delta_0 cos(2\theta)$, in the presence of disorder \cite{Hotta}. The calculation finds excellent agreement with the experimental data as seen in Fig. 2, with 2$\Delta_0=3.8k_BT_c$ and $N_{res}=0.15N(E_F)$. 
 The same model can also consistently explain  the  Knight shift  below $T_c$,  for a choice of $K_{orb}$=0.155\%, but  the difference between the d-wave and the s-wave cases in the Knight shift result is generally small,  which is also true in the present case. In summary, our results in the SC state are compatible  with the photoemission and scanning SQUID experiments that found $d$-wave-like gap in Nd$_{1.85}$Ce$_{0.15}$CuO$_4$ \cite{Tsuei,Sato,Armitage}. 

In conclusion, from $^{63}$Cu NMR measurements in an electron low-doped copper oxide superconductor Pr$_{0.91}$LaCe$_{0.09}$CuO$_{4-y}$,  we find that no pseudogap shows up and that the low-$T$ ($T\leq$ 55 K) spin dynamics is renormalized to well conform to the prediction for the Fermi liquid that persists as the ground state when the superconductivity is removed. This is the first case in which the normal state in the zero-temperature limit was investigated by a microscopic probe.  The NMR data set in the superconducting state, which was obtained for the first time in Re$_{2-x}$Ce$_{x}$CuO$_{4}$,  can be consistently explained by line-node gap model.
These results shed light on understanding  doped Mott insulators and superconductivity derived from them.

We thank H. Kohno, K. Miyake, M. Ogata, Z.-X. Shen, T. Tohyama, C. M. Varma, and T. Xiang for useful discussions, and Y. Okita for assistance in some of the measurements. Support  by  MEXT grants Nos. 11640350, 14540338 (G.-q.Z),  13740216 (M.F) and 12046239 (Y.K and K.Y) is acknowledged.

%\begin{figure}[htbp]
%\begin{center}
%\includegraphics[scale=0.4]{T1T-T_a1.eps}
%\caption{$T$ dependences of $1/T_1T$ of CeCu$_2$(Si$_{0.98}$Ge$_{0.02}$)$_2$ at $P$ = 0, 0.56, and 0.91 GPa and that of Ce$_{0.99}$Cu$_{2.02}$Si$_2$ $P$ = 0.}
%\label{fig:2}
%\end{center}
%\end{figure}

%\begin{figure}[h]
%\begin{center}
%\includegraphics[scale=0.45]{ac-chi_a1.eps}
%\caption{$T$ dependences of ac-$\chi$ of CeCu$_2$(Si$_{0.98}$Ge$_{0.02}$)$_2$ at $P$ = 0, 0.19, 0.56, and 0.91 GPa.}
%\label{fig:2}
%\end{center}
%\end{figure}


\begin{thebibliography}{}
\bibitem{note1}
above $T$=20 K, the anisotropy of $T_1$ is $T$-independent, with $\frac{T_1(H//c)}{T_1(H//a)}$=3.3.
\bibitem{Note}
This $K_{orb,a}$ and  $K_{orb,c}$=0.92\%  are both smaller than the respective values in p-type cuprates (see for example Ref. \cite{Zheng1}). This is consistent with the result of the very small $\nu_Q$ that indicates a reduced hole number in the  Cu3d orbit. Namely, $K_{orb,c}=n_{x^2-y^2} \frac{8\mu_B^2}{E_{xy}-E_{x^2-y^2}}$,  where $n_{x^2-y^2}$ is the hole number in the 3$d_{x^2-y^2}$-orbit and the denominator is the energy difference between the  two designated orbits \cite{Zheng5}.
\end{thebibliography}
\end{document}